\documentclass[aps,prl,showpacs,twocolumn,superscriptaddress,preprintnumbers]{revtex4-1}

\usepackage{graphicx} 
\usepackage{dcolumn}  
\usepackage{longtable}
\usepackage{amsmath}

\usepackage{booktabs}

\usepackage{color}

\renewcommand{\arraystretch}{1.1}

\newcommand{\mev}{\mathrm{MeV}}

\newcommand{\mevm}{\mathrm{MeV}/c^2}
\newcommand{\gev}{\mathrm{GeV}}

\newcommand{\gevm}{\mathrm{GeV}/c^2}

\newcommand{\ee}{e^+e^-}
\newcommand{\uu}{\mu^+\mu^-}
\newcommand{\pp}{\pi^+\pi^-}
\newcommand{\pipi}{\pi\pi}

\newcommand{\U}{\Upsilon}

\newcommand{\Uf}{\Upsilon(5S)}
\newcommand{\Us}{\Upsilon(6S)}
\newcommand{\Uo}{\Upsilon(1S)}

\newcommand{\Ut}{\Upsilon(2S)}

\newcommand{\Un}{\Upsilon(nS)}

\newcommand{\mmpp}{M_{\rm miss}(\pi\pi)}
\newcommand{\mmpi}{M_{\rm miss}(\pi)}
\newcommand{\mmp}{M_{\rm miss}(\pi^+)}
\newcommand{\mmm}{M_{\rm miss}(\pi^-)}

\newcommand{\hb}{h_b(1P)}
\newcommand{\hbp}{h_b(2P)}
\newcommand{\hbn}{h_b(nP)}
\newcommand{\jp}{J/\psi}

\newcommand{\psp}{\psi(2S)}

\newcommand{\pip}{\pi^{+}}
\newcommand{\pipm}{\pi^{\pm}}

\newcommand{\pim}{\pi^{-}}

\newcommand{\fb}{\mathrm{fb}^{-1}}

\newcommand{\etal}{\em et al.}

\newcommand{\ecm}{E_{\rm c.m.}}

\newcommand{\zbo}{Z_b(10610)}
\newcommand{\zbt}{Z_b(10650)}

\begin{document}

\title{\boldmath Energy scan of the $\ee\to\hbn\pp$ $(n=1,2)$ cross
  sections and evidence for $\Upsilon(11020)$ decays into charged
  bottomonium-like states}

\date{\today}

\begin{abstract}
\noindent
Using data collected with the Belle detector at the KEKB
asymmetric-energy $\ee$ collider, we measure the energy dependence of
the $\ee\to\hbn\pp$ $(n=1,2)$ cross sections from thresholds up to
$11.02\,\gev$.  We find clear $\Upsilon(10860)$ and $\Upsilon(11020)$
peaks with little or no continuum contribution. We study the resonant
substructure of the $\Upsilon(11020)\to\hbn\pp$ transitions and find
evidence that they proceed entirely via the intermediate isovector
states $Z_b(10610)$ and $Z_b(10650)$.
The relative fraction of these states is loosely constrained by the
current data: the hypothesis that only $Z_b(10610)$ is produced is
excluded at the level of 3.3 standard deviations, while the hypothesis
that only $Z_b(10650)$ is produced is not excluded at a significant
level.
\end{abstract}


\noaffiliation
\affiliation{Aligarh Muslim University, Aligarh 202002}
\affiliation{University of the Basque Country UPV/EHU, 48080 Bilbao}
\affiliation{Budker Institute of Nuclear Physics SB RAS, Novosibirsk 630090}
\affiliation{Faculty of Mathematics and Physics, Charles University, 121 16 Prague}
\affiliation{University of Cincinnati, Cincinnati, Ohio 45221}
\affiliation{Deutsches Elektronen--Synchrotron, 22607 Hamburg}
\affiliation{University of Florida, Gainesville, Florida 32611}
\affiliation{Justus-Liebig-Universit\"at Gie\ss{}en, 35392 Gie\ss{}en}
\affiliation{SOKENDAI (The Graduate University for Advanced Studies), Hayama 240-0193}
\affiliation{Gyeongsang National University, Chinju 660-701}
\affiliation{Hanyang University, Seoul 133-791}
\affiliation{University of Hawaii, Honolulu, Hawaii 96822}
\affiliation{High Energy Accelerator Research Organization (KEK), Tsukuba 305-0801}
\affiliation{IKERBASQUE, Basque Foundation for Science, 48013 Bilbao}
\affiliation{Indian Institute of Technology Bhubaneswar, Satya Nagar 751007}
\affiliation{Indian Institute of Technology Guwahati, Assam 781039}
\affiliation{Indian Institute of Technology Madras, Chennai 600036}
\affiliation{Indiana University, Bloomington, Indiana 47408}
\affiliation{Institute of High Energy Physics, Chinese Academy of Sciences, Beijing 100049}
\affiliation{Institute of High Energy Physics, Vienna 1050}
\affiliation{INFN - Sezione di Torino, 10125 Torino}
\affiliation{J. Stefan Institute, 1000 Ljubljana}
\affiliation{Kanagawa University, Yokohama 221-8686}
\affiliation{Institut f\"ur Experimentelle Kernphysik, Karlsruher Institut f\"ur Technologie, 76131 Karlsruhe}
\affiliation{King Abdulaziz City for Science and Technology, Riyadh 11442}
\affiliation{Korea Institute of Science and Technology Information, Daejeon 305-806}
\affiliation{Korea University, Seoul 136-713}
\affiliation{Kyoto University, Kyoto 606-8502}
\affiliation{Kyungpook National University, Daegu 702-701}
\affiliation{\'Ecole Polytechnique F\'ed\'erale de Lausanne (EPFL), Lausanne 1015}
\affiliation{P.N. Lebedev Physical Institute of the Russian Academy of Sciences, Moscow 119991}
\affiliation{Faculty of Mathematics and Physics, University of Ljubljana, 1000 Ljubljana}
\affiliation{Ludwig Maximilians University, 80539 Munich}
\affiliation{Luther College, Decorah, Iowa 52101}
\affiliation{University of Maribor, 2000 Maribor}
\affiliation{Max-Planck-Institut f\"ur Physik, 80805 M\"unchen}
\affiliation{School of Physics, University of Melbourne, Victoria 3010}
\affiliation{Middle East Technical University, 06531 Ankara}
\affiliation{University of Miyazaki, Miyazaki 889-2192}
\affiliation{Moscow Physical Engineering Institute, Moscow 115409}
\affiliation{Moscow Institute of Physics and Technology, Moscow Region 141700}
\affiliation{Graduate School of Science, Nagoya University, Nagoya 464-8602}
\affiliation{Nara Women's University, Nara 630-8506}
\affiliation{National Central University, Chung-li 32054}
\affiliation{National United University, Miao Li 36003}
\affiliation{Department of Physics, National Taiwan University, Taipei 10617}
\affiliation{H. Niewodniczanski Institute of Nuclear Physics, Krakow 31-342}
\affiliation{Nippon Dental University, Niigata 951-8580}
\affiliation{Niigata University, Niigata 950-2181}
\affiliation{Novosibirsk State University, Novosibirsk 630090}
\affiliation{Osaka City University, Osaka 558-8585}
\affiliation{Pacific Northwest National Laboratory, Richland, Washington 99352}
\affiliation{University of Pittsburgh, Pittsburgh, Pennsylvania 15260}
\affiliation{University of Science and Technology of China, Hefei 230026}
\affiliation{Seoul National University, Seoul 151-742}
\affiliation{Showa Pharmaceutical University, Tokyo 194-8543}
\affiliation{Soongsil University, Seoul 156-743}
\affiliation{University of South Carolina, Columbia, South Carolina 29208}
\affiliation{Sungkyunkwan University, Suwon 440-746}
\affiliation{School of Physics, University of Sydney, New South Wales 2006}
\affiliation{Department of Physics, Faculty of Science, University of Tabuk, Tabuk 71451}
\affiliation{Tata Institute of Fundamental Research, Mumbai 400005}
\affiliation{Excellence Cluster Universe, Technische Universit\"at M\"unchen, 85748 Garching}
\affiliation{Department of Physics, Technische Universit\"at M\"unchen, 85748 Garching}
\affiliation{Toho University, Funabashi 274-8510}
\affiliation{Department of Physics, Tohoku University, Sendai 980-8578}
\affiliation{Earthquake Research Institute, University of Tokyo, Tokyo 113-0032}
\affiliation{Department of Physics, University of Tokyo, Tokyo 113-0033}
\affiliation{Tokyo Institute of Technology, Tokyo 152-8550}
\affiliation{Tokyo Metropolitan University, Tokyo 192-0397}
\affiliation{University of Torino, 10124 Torino}
\affiliation{Virginia Polytechnic Institute and State University, Blacksburg, Virginia 24061}
\affiliation{Wayne State University, Detroit, Michigan 48202}
\affiliation{Yamagata University, Yamagata 990-8560}
\affiliation{Yonsei University, Seoul 120-749}
  \author{I.~Adachi}\affiliation{High Energy Accelerator Research Organization (KEK), Tsukuba 305-0801}\affiliation{SOKENDAI (The Graduate University for Advanced Studies), Hayama 240-0193} 
  \author{H.~Aihara}\affiliation{Department of Physics, University of Tokyo, Tokyo 113-0033} 
 \author{D.~M.~Asner}\affiliation{Pacific Northwest National Laboratory, Richland, Washington 99352} 
  \author{H.~Atmacan}\affiliation{Middle East Technical University, 06531 Ankara} 
  \author{V.~Aulchenko}\affiliation{Budker Institute of Nuclear Physics SB RAS, Novosibirsk 630090}\affiliation{Novosibirsk State University, Novosibirsk 630090} 
  \author{T.~Aushev}\affiliation{Moscow Institute of Physics and Technology, Moscow Region 141700} 
  \author{R.~Ayad}\affiliation{Department of Physics, Faculty of Science, University of Tabuk, Tabuk 71451} 
  \author{I.~Badhrees}\affiliation{Department of Physics, Faculty of Science, University of Tabuk, Tabuk 71451}\affiliation{King Abdulaziz City for Science and Technology, Riyadh 11442} 
  \author{A.~M.~Bakich}\affiliation{School of Physics, University of Sydney, New South Wales 2006} 
  \author{E.~Barberio}\affiliation{School of Physics, University of Melbourne, Victoria 3010} 
% ...
\collaboration{The Belle Collaboration}

\maketitle

Heavy quarkonium is a bound state of the $c\bar{c}$ or $b\bar{b}$
quarks. The heavy quarks are moving relatively slowly; therefore, a
non-relativistic approximation based on the interaction potential
accurately describes the basic properties of this system~\cite{QWG}.
The first state that did not fit potential model expectations was
observed in 2003 by Belle~\cite{Choi:2003ue}; since then, almost
twenty such states have been reported~\cite{PDG}. They correspond to
high excitations and have masses above the $D\bar{D}$ or $B\bar{B}$
thresholds.

Many quarkonium-like states were found in the energy scans of the
cross sections of $\ee$ annhilation into conventional quarkonia and
light hadrons. Among these are the $Y(4008)$ and $Y(4260)$ in
$\jp\,\pp$~\cite{jppp}, the $Y(4360)$ and $Y(4660)$ in
$\psp\,\pp$~\cite{psppp}, the $\psi(4040)$ and $\psi(4160)$ in
$\jp\,\eta$~\cite{jpeta}, and possibly the $Y(4220)$ in
$h_c\pp$~\cite{hcpp}.
The partial widths of the corresponding transitions are much higher
than expected for conventional quarkonia~\cite{theor}.
Surprisingly, the peaks observed in the cross sections depend on the
final states. In other words, each such charmonium-like state decays
to only one channel with charmonium. To explain this ``selectivity'',
a hadrocharmonium notion is introduced~\cite{hadrocharmonium}: a bound
state of a charmonium and a light hadron. Such a system decays
predominantly into its constituents.

Recent energy scans of the $\ee\to\Un\pp$ $(n=1,2,3)$ cross sections
by Belle~\cite{Chen:2008xia,Santel:2015qga} show that the situation is
different in the sector of bottomonium-like states: all of the cross
sections exhibit peaks of the $\U(10860)$ and $\U(11020)$ resonances
that are also seen in the total hadronic cross section. The observed
decay patterns of $\U(10860)$ and $\U(11020)$ agree with the
expectations for a mixture of the $B^{(*)}_{(s)}\bar{B}^{(*)}_{(s)}$
molecule and conventional bottomonium: open flavor channels dominate,
while channels with quarkonium have anomalously high partial
widths~\cite{voloshin}. The striking difference between
charmonium-like and bottomonium-like states is not yet
understood. Further scans in the bottomonium region are therefore of
high importance. In this Letter, we report the first energy scan of
the $\ee\to\hbn\pp$ $(n=1,2)$ cross sections. We find clear
$\U(10860)$ and $\U(11020)$ peaks without a significant continuum
contribution.

To date, the $\ee\to\hbn\pp$ processes were seen only at a single
energy near the $\U(10860)$ peak~\cite{belle_hb}. They were found to
proceed entirely via the intermediate isovector states $\zbo$ and
$\zbt$ that are situated near the $B\bar{B}^*$ and $B^*\bar{B}^*$
thresholds~\cite{belle_zb} and likely have corresponding molecular
structures~\cite{Bondar:2011ev}. Here, we report on the resonant
substructure study of the $\U(11020)\to\hbn\pp$ decays, where we find
first evidence for intermediate $Z_b$ states.
Hereinafter, the $\U(10860)$ and $\U(11020)$ are referred to, for
brevity, as the $\Uf$ and $\Us$ according to the potential model
assignment.

We use $121.4\,\fb$ of on-resonance $\Uf$ data taken at three energies
close to $10.866\,\gev$, as well as $1\,\fb$ of data taken at each of
19 different energies between $10.77$ and $11.02\,\gev$. These data
were collected with the Belle detector~\cite{BELLE_DETECTOR} at the
KEKB asymmetric-energy $\ee$ collider~\cite{KEKB}.

The $\ee\to\hbn\pp$ processes are reconstructed inclusively based on
the $\pp$ missing mass,
$\mmpp=\sqrt{(\ecm-E_{\pipi}^*)^2-p_{\pipi}^{*2}}$, where $\ecm$ is
the center-of-mass (c.m.) energy and $E^*_{\pipi}$ and $p^*_{\pipi}$
are the energy and momentum of the $\pp$ pair as measured in the
c.m.\ frame.
The c.m.\ energy is calibrated using the
$\ee\to\Un\pp\to\uu\pp$ and $\ee\to\uu$ processes, as described in
Ref.~\cite{Santel:2015qga}.
This analysis closely follows previous Belle
publications~\cite{belle_hb,belle_zb,Mizuk:2012pb}. 

We use a general hadronic event selection with requirements on the
position of the primary vertex, track multiplicity, and the total
energy and momentum of the event~\cite{hadronbj}. These criteria
suppress Bhabha, $\uu$, $\tau^+\tau^-$, two-photon and beam--gas
processes.
Continuum $\ee\to q\bar{q}$ ($q=u$, $d$, $s$, $c$) events have
jet-like shapes in contrast to the spherically symmetric signal events
and are suppressed by a requirement on the ratio of the second to
zeroth Fox-Wolfram moments: $R_2<0.3$~\cite{Fox-Wolfram}.
We only consider positively identified $\pp$ candidates that originate
from the interaction point region.

The measurements of the cross sections are performed with an
additional requirement on the \textit{single-pion} $\pipm$ missing
mass,
\begin{equation}
  10.59\,\gevm<M_{\mathrm{miss}}(\pipm)<10.67\,\gevm,
  \label{zb_cut}
\end{equation}
which selects signal events proceeding via the intermediate $\zbo$ or
$\zbt$ states.
We combine the $\mmpp$ distribution for $\pip$
satisfying~(\ref{zb_cut}) and that for $\pim$
satisfying~(\ref{zb_cut}).
The $\pp$ pairs with both $\mmp$ and $\mmm$ in the $Z_b$ mass window
are counted twice. (If they are counted only once, the combinatorial
background develops a dip slightly above the $\hbp$ signal, making the
background parameterization difficult.) We take the double entries
into account by correcting the errors of the $\mmpp$ histogram and,
based on Monte Carlo (MC) simulation, the $\hbp$ signal yields.

We fit the $\mmpp$ distribution in the $\hb$ and $\hbp$ intervals,
defined as $9.8\,\gevm-10.0\,\gevm$ and $10.17\,\gevm-10.34\,\gevm$,
respectively.
The fit function is the sum of the $\hbn$ signal and combinatorial-
and peaking-background components.
The shapes of the $\hbn$ signals are determined by convolving the
probability density of the initial state radiation (ISR) process with
the experimental resolution, described by a Gaussian. We use the ISR
probability, calculated up to the second order~\cite{Kuraev:1985hb},
and take into account the energy dependence of the $\ee\to\hbn\pp$
cross sections using an iterative procedure. The resolution is
determined using the exclusively reconstructed decays $\Uf\to\Un\pp$,
$\Un\to\uu$ to be $(6.84\pm0.13)\,\mevm$ for the $\hb$ and
$(6.15\pm0.22)\,\mevm$ for the $\hbp$. The resolution is dominated by
c.m.\ energy smearing.
The $\hbn$ masses are fixed at the previous Belle
measurement~\cite{Mizuk:2012pb}.
We normalize the signal density functions in such a way that the
measured $\hbn$ yields include the ISR correction,
$1+\delta_{\mathrm{ISR}}$, and can be used directly to measure the
Born cross sections.
The combinatorial background is described by a fourth-order Chebyshev
polynomial in both fit intervals. The order is chosen by maximizing
the confidence level of the fit.

Using MC simulation, we find that combining a random pion that
satisfies the $Z_b$ mass requirement and a signal pion from
$Z_b\to\hbn\pi$ produces a broad bump under the $\hbn$ signal. This
background is incorporated within the combinatorial background and
results in minor corrections in the $\hb$ and $\hbp$ yields of
$0.99\pm0.01$ and $0.995\pm0.005$, respectively.
The $\pp$ pairs originating from the $\Ut\to\Uo\pp$ transitions with
the $\Ut$ produced inclusively or via ISR result in a peak at
$\ecm-[m_{\Ut}-m_{\Uo}]$ that is inside the $\hbp$ fit interval for
the c.m.\ energies close to the $\Uf$. The shape of this peaking
background is found to be a Gaussian with $\sigma=11\,\mevm$. Its
normalization is floated in the fit.


To determine the reconstruction efficiency, we use
phase-space-generated MC, weighted in $\mmpi$ according to the fit
results for the $\Uf\to\hb\pp$ transitions~\cite{belle_zb} and in
angular variables according to the expectations for the $Z_b$
spin-parity $J^P=1^+$~\cite{Garmash:2014dhx}. The efficiencies for the
$\hb\pp$ and $\hbp\pp$ channels are in the range $40-55\%$ and
$35-50\%$, respectively; they rise with c.m.\ energy.
At the lowest energy point, there is a drop of efficiency by a factor
of two since this point is close to the kinematic boundary and the
pion momenta are low.

\begin{figure}[tb]
\includegraphics[width=0.48\textwidth]{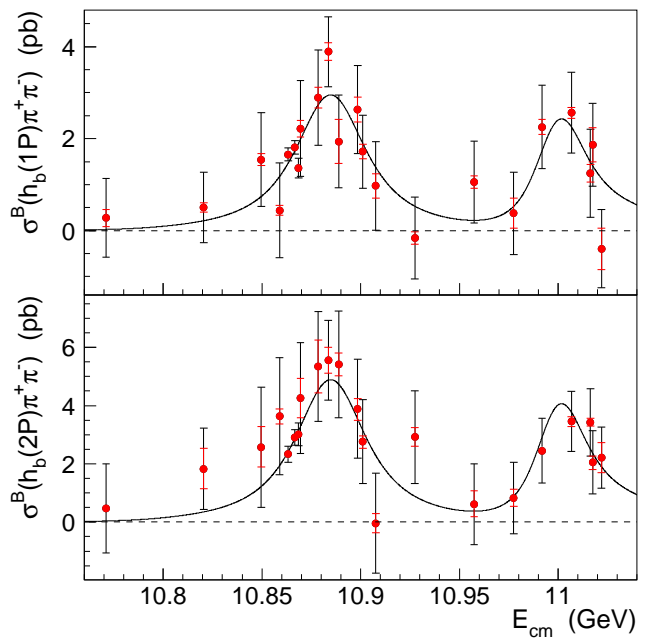}
\caption{ (colored online) The cross sections for the $\ee\to\hb\pp$
  (top) and $\ee\to\hbp\pp$ (bottom) as functions of
  c.m.\ energy. Points with error bars are the data; outer error bars
  indicate statistical uncertainties and inner red error bars indicate
  uncorrelated systematic uncertainties. The solid curves are the fit
  results. }
\label{hbpp_born_simfit}
\end{figure}

At each energy, the Born cross section is determined according to the
formula:
\begin{equation}
\sigma^B(\ee\to\hbn\pp)=\frac{N}{L\; \varepsilon\; |1-\Pi|^2},
\end{equation}
where $N$ is the number of signal events determined from the $\mmpp$
fit that includes the ISR correction, $L$ is the integrated
luminosity, $\varepsilon$ is the reconstruction efficiency and
$|1-\Pi|^2$ is the vacuum polarization correction~\cite{Actis:2010gg},
which is in the range $0.927-0.930$.
The resulting cross sections are shown in
Fig.~\ref{hbpp_born_simfit}. 
The cross sections, averaged over the three high statistics
on-resonance points at $\ecm=(10865.6\pm2.0)\,\mev$, are
\begin{align}
\label{eq:born1_onres}
\sigma^B(\ee\to\hb\pp) & =1.66\pm0.09\pm0.10\,\mathrm{pb},\\
\label{eq:born2_onres}
\sigma^B(\ee\to\hbp\pp)& =2.70\pm0.17\pm0.19\,\mathrm{pb}. 
\end{align}
The ratio of the cross sections is $0.616\pm0.052\pm0.017$.
Here and elsewhere in this Letter, the first uncertainties are
statistical and the second are systematic.

The systematic uncertainties in the signal yields originate from the
signal and background shapes. The relative uncertainty due to the
$\mmpp$ resolution is correlated among different energy points and is
equal to $1.4\%$ for the $\hb$ and $3.3\%$ for the $\hbp$. The
uncertainties due to the $\hbn$ masses and ISR tail shapes are found
to be negligible. To estimate the background-shape contribution, we
vary the fit interval limits by about $50\,\mev$ and the polynomial
order for each fit interval.
%
%
The corresponding uncertainties are considered uncorrelated and are
1.1\% and 2.5\% for the on-resonance cross sections in
Eqs.~(\ref{eq:born1_onres}) and (\ref{eq:born2_onres}), respectively.

A relative uncertainty in the efficiency contributes to the correlated
systematic uncertainty. An uncertainty due to the $Z_b$ mass
requirement of $^{+1.0}_{-1.8}\%$ is estimated by varying the $Z_b$
parameters by $\pm1\sigma$ and taking into account correlations among
different parameters. The efficiency of the $R_2$ requirement is
studied using inclusively reconstructed $\Uf\to\U(nS)\pp$ decays. We
find good agreement between data and MC and assign the $5\%$
statistical uncertainty in data as a systematic uncertainty due to the
$R_2$ requirement. Finally, we assign a $1\%$ uncertainty per track
due to possible differences in the reconstruction efficiency between
data and MC.

An uncertainty in the luminosity of 1.4\% is primarily due to the
simulation of Bhabha scattering that is used for its determination and
is correlated among energy points. We add in quadrature all the
contributions to find the total systematic uncertainties shown in
Eqs.~(\ref{eq:born1_onres}) and (\ref{eq:born2_onres}).
The values of the cross sections for all energy points are provided
in Ref.~\cite{suppl}.

The shapes of the $\hb\pp$ and $\hbp\pp$ cross sections look very
similar. They show clear $\Uf$ and $\Us$ peaks without significant
continuum contributions.
We perform a simultaneous fit of the shapes, adding in quadrature the
statistical and uncorrelated systematic uncertainties at each energy
point.
We use the coherent sum of two Breit-Wigner amplitudes:
\begin{equation}
A_n\;\Phi_n(s)\;|F_{\rm BW}(s,M_5,\Gamma_5)
+a\,e^{i\,\phi}\,F_{\rm BW}(s,M_6,\Gamma_6)|^2,
\label{eq:fit_fun}
\end{equation}
where $s\equiv\ecm^{\,2}$, $\Phi_n(s)$ is the phase space
calculated numerically, taking into account the measured $Z_b$ line
shape~\cite{belle_zb}, and $F_{\rm BW}(s,M,\Gamma) =
M\Gamma/(s-M^2+iM\Gamma)$ is a Breit-Wigner amplitude. The fit
parameters $M_5$, $\Gamma_5$, $M_6$, $\Gamma_6$, $a$ and $\phi$ are
common for the two channels, while only the normalization coefficients
$A_n$ are different.
Equation (\ref{eq:fit_fun}) is convolved with the $\ecm$ resolution of
$(5.0\pm0.4)\,\mev$, which is found using exclusively reconstructed
$\Uf\to\Un\pp$ events.
The fitted functions are shown in Fig.~\ref{hbpp_born_simfit}. The
confidence level of the fit is 93\%.
The fit results are:
\begin{align}
M_5 &=(10884.7^{+3.6}_{-3.4}{^{+8.9}_{-1.0}})\,\mevm, \\
\Gamma_5 &=(40.6^{+12.7}_{-\phantom{1}8.0}{^{+\phantom{1}1.1}_{-19.1}})\,\mev, \\ 
M_6 &=(10999.0^{+7.3}_{-7.8}{^{+16.9}_{-\phantom{1}1.0}})\,\mevm, \\
\Gamma_6 &=(27^{+27}_{-11}{^{+\phantom{1}5}_{-12}})\,\mev, \\
a &=0.65^{+0.36}_{-0.12}{^{+0.17}_{-0.10}} \;\; \mathrm{and} \;\;
\phi=(0.1^{+0.4}_{-0.8}\pm0.3)\,\pi.
\end{align}
The measured masses and widths agree with the results of the $\Un\pp$
scan~\cite{Santel:2015qga}.

The first error in the fit results is not purely statistical but
includes uncorrelated systematic uncertainties in the cross
sections. The contributions of other considered sources are listed in
Table~\ref{tab:syst}.
\begin{table}[htbp]
\caption{ The systematic uncertainties in the $\Uf$ and $\Us$ masses
  (in $\mevm$), widths (in $\mev$), amplitude $a$, and phase $\phi$
  (in units of $\pi$). }
\label{tab:syst}
\renewcommand{\arraystretch}{1.3}
\begin{tabular}{@{}l@{\hspace{3mm}}c@{\hspace{3mm}}c@{\hspace{3mm}}c@{\hspace{3mm}}c@{\hspace{3mm}}c@{\hspace{3mm}}c@{}}
\hline
& $M_5$ & $\Gamma_5$ & $M_6$ & $\Gamma_6$ & $a$ & $\phi$ \\
\hline
Fit model          & $^{+8.9}_{-0.1}$ & $^{+\phantom{1}0.4}_{-19.1}$ & $^{+16.7}_{-\phantom{1}0.0}$ & $^{+\phantom{1}0.0}_{-11.5}$ & $^{+0.12}_{-0.00}$ & $^{+0.09}_{-0.00}$ \\
$Z_b$ substructure & $^{+0.2}_{-0.0}$ & $^{+0.0}_{-0.2}$ & $^{+0.1}_{-0.0}$ & $^{+0.7}_{-0.0}$ & $^{+0.11}_{-0.00}$ & $^{+0.00}_{-0.29}$ \\
$\sqrt{s}$ scale   & $1.0$ & $1.0$ & $^{+3.0}_{-1.0}$ & $^{+4.7}_{-1.0}$ & $^{+0.00}_{-0.10}$ & $^{+0.25}_{-0.00}$ \\
Resolution         & $0.0$ & $^{+0.3}_{-0.2}$ & $0.1$ & $0.6$ & $0.0$ & $^{+0.01}_{-0.00}$ \\
\hline
Total              & $^{+8.9}_{-1.0}$ & $^{+\phantom{1}1.1}_{-19.1}$ & $^{+16.9}_{-\phantom{1}1.0}$ & $^{+\phantom{1}4.8}_{-11.5}$ & $^{+0.17}_{-0.10}$ & $^{+0.27}_{-0.29}$ \\
\hline
\end{tabular}
\end{table}

To study systematic uncertainties due to the fit model, we introduce a
non-resonant continuum amplitude, $b\,e^{i\,\delta}$. The significance
of this contribution is only $1.6\sigma$. However, the shifts in the
fit results are large, and this is the dominant source of systematic
uncertainty.
We also consider the possibility that the parameters $a$ and $\phi$
are different in the $\hb\pp$ and $\hbp\pp$ channels. We find that the
values in the two channels agree and the shifts in masses and widths
are small.
Using MC pseudo-experiments, we find that there is no significant fit
bias.

If the resonant substructures of the $\Uf$ and $\Us$ decays are
different, the $\Uf$ and $\Us$ amplitudes in~(\ref{eq:fit_fun})
are not fully coherent, and the interference term is suppressed by a
decoherence factor $k$~\cite{Santel:2015qga}. If only $\zbo$ is
produced at the $\Us$, $k$ is calculated numerically to be $0.62$; if
only $\zbt$ is produced, $k$ is $0.80$. We introduce these factors in
the fit and take into account that the efficiency of the $Z_b$ mass
requirement is smaller for a single $Z_b$ state compared to two $Z_b$
states by 12\% since the two $Z_b$ states interfere destructively
outside their signal region.

We account for an uncertainty in the $\ecm$ scale and the
uncertainty in the $\ecm$ resolution.
We add in quadrature the contributions of the various sources to
determine the total systematic uncertainties.

\begin{figure}[htbp]
\includegraphics[width=0.46\textwidth]{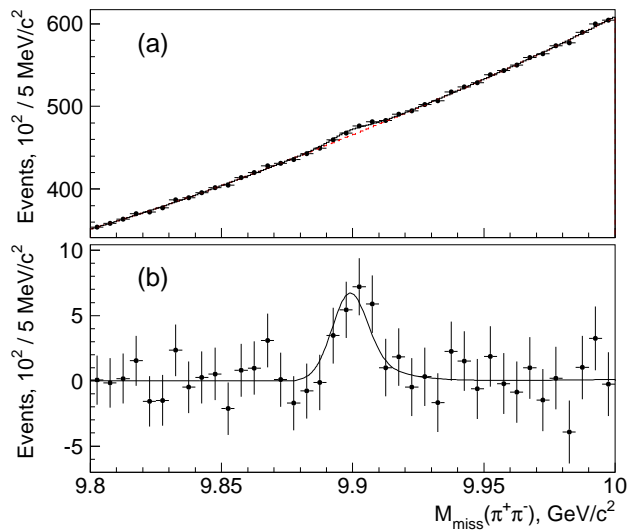}
\caption{ (colored online) The $\mmpp$ spectrum in the $\hb$ region
  for the combined data samples of five energy points near the
  $\Us$. In (a) the data are the points with error bars with the fit
  function (solid curve) and background (red dashed curve)
  overlaid. (b) shows the background-subtracted data (points with
  error bars) with the signal component of the fit overlaid (solid
  curve). }
\label{mmpp_hb1}
\end{figure}

\begin{figure}[htbp]
\includegraphics[width=0.46\textwidth]{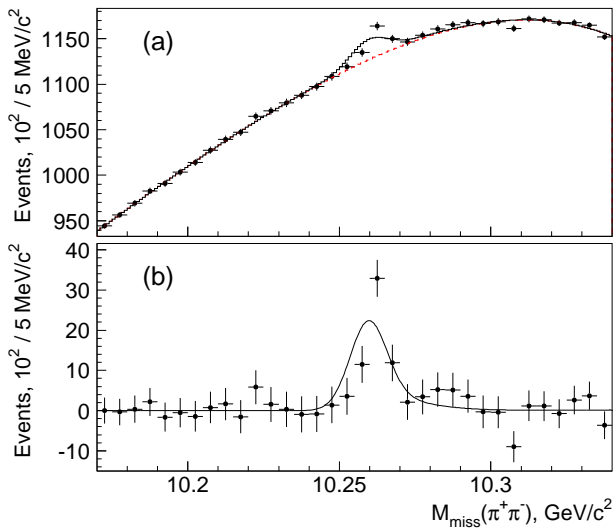}
\caption{ (colored online) The $\mmpp$ spectrum in the $\hbp$ interval
  for the combined data samples of five energy points near the
  $\Us$. The legend is the same as in Fig.~\ref{mmpp_hb1}. }
\label{mmpp_hb2}
\end{figure}

To study the resonant substructure of the $\Us\to\hbn\pp$ transitions,
we combine the data samples of the five highest-energy points. The
fits to the corresponding $\mmpp$ spectra in the $\hb$ and $\hbp$
intervals are shown in Figs.~\ref{mmpp_hb1} and \ref{mmpp_hb2}.
The $h_b(nP)$ signal density functions are determined by averaging
over the data samples that are combined; we use weights proportional
to the integrated luminosity and the cross section at each energy.
The confidence levels of the fits are 50\% and 52\%,
respectively. From Wilks' theorem~\cite{wilks}, we find that the
significances of the $\hb$ and $\hbp$ signals are $3.5\sigma$ and
$5.3\sigma$, respectively, including systematic uncertainty,
determined by varying the polynomial order. Thus, we find the first
evidence for the $\Us\to\hb\pp$ transition and observe for the first
time the $\Us\to\hbp\pp$ transition.

\begin{figure}[htbp]
\includegraphics[height=5cm]{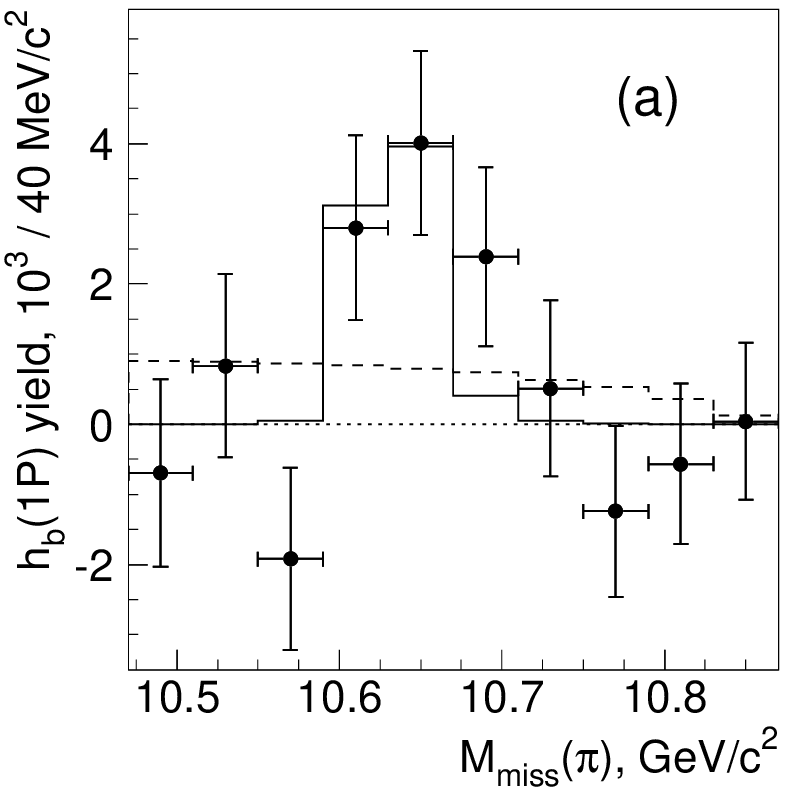}\hfill
\includegraphics[height=5cm]{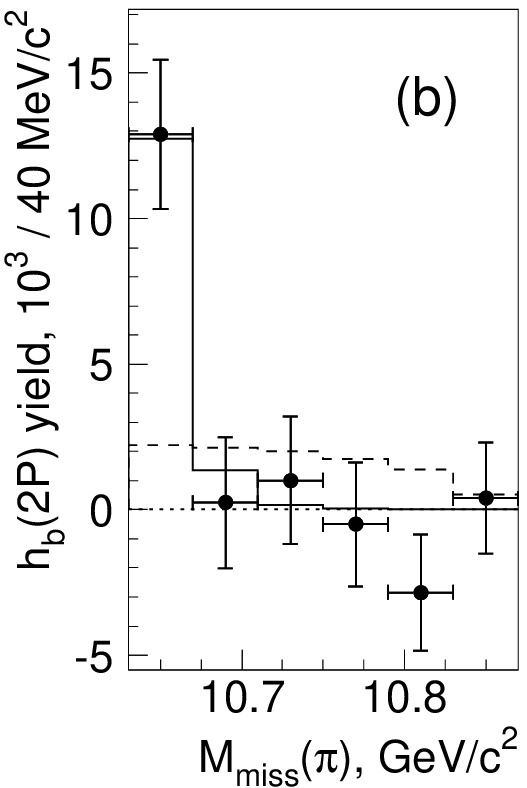}
\caption{ The efficiency-corrected yields of $\hb\pp$ (a) and
  $\hbp\pp$ (b) as functions of $\mmpi$ for the combined data samples
  of five energy points in the $\Us$ region. Points represent data;
  the solid histogram represents the fit result with the $Z_b$ signal
  shape fixed from the $\Uf$ analysis; the dashed histogram represents
  the result of the fit with a phase space distribution. }
\label{hb_vs_mmp_fits}
\end{figure}

We release the requirement of an intermediate $Z_b$ and fit the
$\mmpp$ spectra in bins of $\mmpi$ to measure the $\hbn\pp$ yields as
functions of $\mmpi$.
The distribution of the phase-space-generated signal events in the
$\mmp$ vs. $\mmm$ plane has the shape of a narrow slanted band; each
structure at high values of $M_{\rm miss}(\pi^{\pm})$ produces a
``reflection'' at small values of $M_{\rm miss}(\pi^{\mp})$. We
combine the $\mmpp$ spectra for the corresponding $\mmp$ and $\mmm$
bins and consider the upper half of the available $\mmpi$
range. Thereby, we consider all signal events and avoid double
counting.
The yields, corrected for the reconstruction efficiencies, are shown
in Fig.~\ref{hb_vs_mmp_fits}.
The data are not distributed uniformly in phase space; they populate
the $\zbo$ and $\zbt$ mass region.
We fit the data to a shape where the $\zbo$ and $\zbt$ parameters are
fixed to the $\Uf\to Z_b\pi\to\hb\pp$ result and the non-resonant
contribution is set to zero~\cite{belle_zb}. Such a model describes
the data well: the confidence levels of the fits are 65\% and 77\% for
the $\hb$ and $\hbp$, respectively. The phase space hypothesis is
excluded relative to this model at the $3.6\sigma$ and $4.5\sigma$
levels in the $\hb\pp$ and $\hbp\pp$ channels, respectively. The
single $\zbo$ hypothesis is excluded at the $3.3\sigma$ level in the
$\hb\pp$ channel, while the single $\zbt$ hypothesis cannot be
excluded at a significant level.
In the $\hbp\pp$ channel, the $\zbo^{\pm}$ and $\zbt^{\pm}$ signals
overlap with the $\zbt^{\mp}$ and $\zbo^{\mp}$ reflections,
respectively, which obscures the determination of the relative yields.
The exclusion levels are determined using pseudo-experiments from the
$\chi^2$ differences of the two hypotheses being compared, and include
systematic uncertainty.

In conclusion, we have measured the energy dependence of the
$\ee\to\hbn\pp$ ($n=1,2$) cross sections. We find two peaks
corresponding to the $\Uf$ and $\Us$ states and measure their
parameters, which agree with the results from
Ref.~\cite{Santel:2015qga}. The data are consistent with no continuum
contribution.

We report first evidence for $\Us\to\hb\pp$ and first observation of
the $\Us\to\hbp\pp$ transitions. We study their resonant substructures
and find evidence that they proceed entirely via the intermediate
isovector states $\zbo$ and $\zbt$.
Their relative fraction is loosely constrained by the current data:
the hypothesis that only $Z_b(10610)$ is produced is excluded at the
$3.3\sigma$ level, while the hypothesis that only $Z_b(10650)$ is
produced is not excluded at a significant level.

The shapes of the $\ee\to\hbn\pp$ and $\ee\to\Un\pp$ cross sections
look similar. The only significant difference is a smaller relative
yield of $\Un\pp$ at the $\Us$. Since the $\hbn\pp$ final states are
produced only via intermediate $Z_b$ while $\Un\pp$ at the $\Uf$ are
produced both via $Z_b$ and non-resonantly, this difference indicates
that the non-resonant contributions in $\Un\pp$ are suppressed at the
$\Us$.


We thank the KEKB group for excellent operation of the accelerator;
the KEK cryogenics group for efficient solenoid operations; and the
KEK computer group, the NII, and PNNL/EMSL for valuable computing and
SINET4 network support.  We acknowledge support from MEXT, JSPS and
Nagoya's TLPRC (Japan); ARC (Australia); FWF (Austria); NSFC and CCEPP
(China); MSMT (Czechia); CZF, DFG, EXC153, and VS (Germany); DST
(India); INFN (Italy); MOE, MSIP, NRF, BK21Plus, WCU and RSRI (Korea);
MNiSW and NCN (Poland); MES, RFAAE and RSF under Grant No. 15-12-30014
(Russia); ARRS (Slovenia); IKERBASQUE and UPV/EHU (Spain); SNSF
(Switzerland); MOE and MOST (Taiwan); and DOE and NSF (USA).

\end{document}


\centerline{\bf \large Supplemental Material}\vspace{0.8cm}

The $\ee\to\hbn\pp$ $(n=1,2)$ Born cross sections for all energy
points are presented in Table~\ref{tab:cross_sect}.
\begin{table*}[tbhp]
\caption{ Center-of-mass energies, integrated luminosities and Born
  cross sections for all energy points. The first uncertainty in the
  energy is uncorrelated, the second is correlated. The three
  uncertainties in the cross sections are statistical, uncorrelated
  systematic and correlated systematic. }
\label{tab:cross_sect}
\renewcommand{\arraystretch}{1.2}
\begin{tabular}{@{}c@{\hspace{4mm}}c@{\hspace{4mm}}c@{\hspace{4mm}}r@{\hspace{4mm}}r@{}} 
\hline
\# & $\sqrt{s}$, $\mev$ & Luminosity, $\fb$ &
$\sigma(\ee\to\hb\pp)$, pb & $\sigma(\ee\to\hbp\pp)$, pb \\ 
\hline 
 1 & $11022.0^{+0.4}_{-5.3}\pm1.0$ & 0.98 & $-0.39\pm0.85\pm0.45\pm0.02$ & $2.21\pm1.05\pm0.51\pm0.15$ \\
 2 & $11017.5\pm4.0\pm1.0$ & 0.86 & $1.87\pm0.90\pm0.37\pm0.11$ & $2.05\pm1.09^{+0.21}_{-0.24}\pm0.14$ \\
 3 & $11016.4^{+0.4}_{-4.6}\pm1.0$ & 0.77 & $1.25\pm0.96\pm0.19\pm0.08$ & $3.42\pm1.15^{+0.15}_{-0.13}\pm0.23$ \\
 4 & $11006.8^{+0.4}_{-3.9}\pm1.0$ & 0.98 & $2.57\pm0.88\pm0.13\pm0.15$ & $3.45\pm1.03\pm0.17\pm0.23$ \\
 5 & $10991.9\pm0.4\pm1.0$ & 0.99 & $2.25\pm0.91\pm0.16\pm0.14$ & $2.45\pm1.11\pm0.13\pm0.17$ \\
 6 & $10977.5\pm0.4\pm1.0$ & 1.00 & $0.38\pm0.90\pm0.33\pm0.02$ & $0.83\pm1.23\pm0.29\pm0.06$ \\
 7 & $10957.5\pm4.0\pm1.0$ & 0.97 & $1.05\pm0.89\pm0.14\pm0.06$ & $0.60\pm1.39^{+0.47}_{-0.42}\pm0.04$ \\
 8 & $10927.5\pm4.0\pm1.0$ & 1.15 & $-0.16\pm0.89\pm0.13\pm0.01$ & $2.92\pm1.59\pm0.33\pm0.20$ \\
 9 & $10907.7^{+0.4}_{-4.9}\pm1.0$ & 0.98 & $0.97\pm0.96\pm0.27\pm0.06$ & $-0.04\pm1.72\pm0.33\pm0.00$ \\
 10 & $10901.1^{+1.1}_{-4.9}\pm1.0$ & 1.42 & $1.72\pm0.79\pm0.16\pm0.10$ & $2.76\pm1.44^{+0.20}_{-0.24}\pm0.19$ \\
 11 & $10898.5^{+0.4}_{-4.0}\pm1.0$ & 0.98 & $2.63\pm0.96\pm0.27\pm0.16$ & $3.89\pm1.70^{+0.35}_{-0.40}\pm0.26$ \\
 12 & $10888.9^{+0.4}_{-2.0}\pm1.0$ & 0.99 & $1.94\pm1.00\pm0.48\pm0.12$ & $5.41\pm1.83\pm0.40\pm0.37$ \\
 13 & $10883.6^{+0.9}_{-2.1}\pm1.0$ & 1.85 & $3.89\pm0.76\pm0.19\pm0.23$ & $5.55\pm1.37\pm0.45\pm0.38$ \\
 14 & $10878.5^{+0.4}_{-1.4}\pm1.0$ & 0.98 & $2.89\pm1.04\pm0.23\pm0.17$ & $5.34\pm1.89\pm0.91\pm0.36$ \\
 15 & $10869.5^{+0.4}_{-2.0}\pm1.0$ & 0.98 & $2.22\pm1.04\pm0.18\pm0.13$ & $4.26\pm1.90\pm0.68\pm0.29$ \\
 16 & $10868.6\pm0.2\pm0.5$ & 22.94 & $1.36\pm0.21\pm0.04\pm0.08$ & $3.01\pm0.39\pm0.08\pm0.20$ \\
 17 & $10866.7\pm0.2\pm0.5$ & 50.47 & $1.81\pm0.15\pm0.04\pm0.11$ & $2.91\pm0.26\pm0.08\pm0.20$ \\
 18 & $10863.3\pm0.2\pm0.5$ & 47.65 & $1.66\pm0.15\pm0.03\pm0.10$ & $2.33\pm0.27\pm0.11\pm0.16$ \\
 19 & $10858.9^{+0.4}_{-2.0}\pm1.0$ & 0.99 & $0.44\pm1.03\pm0.11\pm0.03$ & $3.63\pm2.01^{+0.25}_{-0.27}\pm0.25$ \\
 20 & $10849.7^{+0.4}_{-1.2}\pm1.0$ & 0.99 & $1.54\pm1.02\pm0.13\pm0.09$ & $2.57\pm2.07^{+0.71}_{-0.68}\pm0.17$ \\
 21 & $10820.5^{+6.5}_{-0.4}\pm1.0$ & 1.70 & $0.50\pm0.77\pm0.10\pm0.03$ & $1.83\pm1.40\pm0.70\pm0.12$ \\
 22 & $10771.1\pm1.8\pm1.0$ & 0.95 & $0.28\pm0.86\pm0.19\pm0.02$ & $0.47\pm1.52\pm0.12\pm0.03$ \\
\hline
\end{tabular}
\end{table*}